\documentclass[jap,aip,reprint,amssymb,amsmath,numerical,superscriptaddress]{revtex4-1}
\usepackage{graphicx}
\usepackage{dcolumn}
\usepackage{bm}
\usepackage{color}
\usepackage{amssymb}
\usepackage{amsmath}
\usepackage{epstopdf}
\usepackage{bbding}
\usepackage[mathlines]{lineno}
\usepackage[dvipdfm, pdfstartview=FitH, CJKbookmarks=true, bookmarksnumbered=true, bookmarksopen=true, colorlinks, pdfborder=001, linkcolor=black, anchorcolor=blue, citecolor=blue]{hyperref}
\usepackage{lastpage}
\usepackage{fancyhdr}
\begin{document}
\preprint{Pt/IrMn/YIG preprint}
\title{Investigation of anomalous-Hall and spin-Hall effects of antiferromagnetic IrMn sandwiched by Pt and YIG layers}
\author{T. Shang}
\thanks{Present address: Swiss light source $\&$ Laboratory for Scientific Developments and Novel Materials, Paul Scherrer Institut, CH-5232 Villigen PSI, Switzerland}
\affiliation{Key Laboratory of Magnetic Materials and Devices $\&$ Zhejiang Province Key Laboratory of Magnetic Materials and Application
Technology, Ningbo Institute of Material Technology and Engineering, Chinese Academy of Sciences, Ningbo, Zhejiang 315201, China}
\author{H. L. Yang}
\affiliation{Key Laboratory of Magnetic Materials and Devices $\&$ Zhejiang Province Key Laboratory of Magnetic Materials and Application
Technology, Ningbo Institute of Material Technology and Engineering, Chinese Academy of Sciences, Ningbo, Zhejiang 315201, China}
\author{Q. F. Zhan}
\email{zhanqf@nimte.ac.cn}
\affiliation{Key Laboratory of Magnetic Materials and Devices $\&$ Zhejiang Province Key Laboratory of Magnetic Materials and Application
Technology, Ningbo Institute of Material Technology and Engineering, Chinese Academy of Sciences, Ningbo, Zhejiang 315201, China}
\author{Z. H. Zuo}
\affiliation{Key Laboratory of Magnetic Materials and Devices $\&$ Zhejiang Province Key Laboratory of Magnetic Materials and Application
Technology, Ningbo Institute of Material Technology and Engineering, Chinese Academy of Sciences, Ningbo, Zhejiang 315201, China}
\author{Y. L. Xie}
\affiliation{Key Laboratory of Magnetic Materials and Devices $\&$ Zhejiang Province Key Laboratory of Magnetic Materials and Application
Technology, Ningbo Institute of Material Technology and Engineering, Chinese Academy of Sciences, Ningbo, Zhejiang 315201, China}
\author{L. P. Liu}
\affiliation{Key Laboratory of Magnetic Materials and Devices $\&$ Zhejiang Province Key Laboratory of Magnetic Materials and Application
Technology, Ningbo Institute of Material Technology and Engineering, Chinese Academy of Sciences, Ningbo, Zhejiang 315201, China}
\author{S. L. Zhang}
\affiliation{Key Laboratory of Magnetic Materials and Devices $\&$ Zhejiang Province Key Laboratory of Magnetic Materials and Application
Technology, Ningbo Institute of Material Technology and Engineering, Chinese Academy of Sciences, Ningbo, Zhejiang 315201, China}
\author{Y. Zhang}
\affiliation{Key Laboratory of Magnetic Materials and Devices $\&$ Zhejiang Province Key Laboratory of Magnetic Materials and Application
Technology, Ningbo Institute of Material Technology and Engineering, Chinese Academy of Sciences, Ningbo, Zhejiang 315201, China}
\author{H. H. Li}
\affiliation{Key Laboratory of Magnetic Materials and Devices $\&$ Zhejiang Province Key Laboratory of Magnetic Materials and Application
Technology, Ningbo Institute of Material Technology and Engineering, Chinese Academy of Sciences, Ningbo, Zhejiang 315201, China}
\author{B. M. Wang}
\affiliation{Key Laboratory of Magnetic Materials and Devices $\&$ Zhejiang Province Key Laboratory of Magnetic Materials and Application
Technology, Ningbo Institute of Material Technology and Engineering, Chinese Academy of Sciences, Ningbo, Zhejiang 315201, China}
\author{Y. H. Wu}
\affiliation{Department of Electrical and Computer Engineering, National University of Singapore, 4 Engineering Drive 3 117583, Singapore}
\author{S. Zhang}
\email{zhangshu@email.arizona.edu}
\affiliation{Department of Physics, University of Arizona, Tucson, Arizona 85721, USA}
\author{Run-Wei Li}
\email{runweili@nimte.ac.cn}
\affiliation{Key Laboratory of Magnetic Materials and Devices $\&$ Zhejiang Province Key Laboratory of Magnetic Materials and Application
Technology, Ningbo Institute of Material Technology and Engineering, Chinese Academy of Sciences, Ningbo, Zhejiang 315201, China}
\date{\today}

\begin{abstract}
We report an investigation of temperature and IrMn layered thickness dependence of anomalous-Hall resistance (AHR), anisotropic magnetoresistance (AMR), and magnetization on Pt/Ir$_{20}$Mn$_{80}$/Y$_3$Fe$_5$O$_{12}$ (Pt/IrMn/YIG) heterostructures. The magnitude of AHR is dramatically enhanced compared with Pt/YIG bilayers. The enhancement is much more profound at higher temperatures and peaks at the IrMn thickness of 3 nm. The observed spin-Hall magnetoresistance (SMR) in the temperature range of 10-300 K indicates that the spin current generated in the Pt layer can penetrate the entire thickness of the IrMn layer to interact with the YIG layer. The lack of conventional anisotropic magnetoresistance (CAMR) implies that the insertion of the IrMn layer between Pt and YIG efficiently suppresses the magnetic proximity effect (MPE) on induced Pt moments by YIG. Our results suggest that the dual roles of the IrMn insertion in Pt/IrMn/YIG heterostructures are to block the MPE and to transport the spin current between Pt and YIG layers. We discuss possible mechanisms for the enhanced AHR.
\end{abstract}
\maketitle
\section{INTRODUCTION}
Antiferromagnts (AFMs) are promising candidates for spintronic applications.~\cite{macdonald2011} Compared to ferromagnetic (FM) materials, the AFMs exhibit unique advantages, e.g., zero net magnetization, insensitivity to the external magnetic perturbation, lack of stray field, and access to extremely high frequency. Recently, the generation and transmission of spin current in AFMs have attracted great attention. The spin pumping studies on (Pt, Ta)/(NiO, CoO)/Y$_3$Fe$_5$O$_{12}$ (YIG) heterostructures demonstrate that the spin current generated in YIG layer can pass through the antiferromagnetic (AFM) insulator NiO or CoO layer and can be detected in Pt or Ta layer by inverse spin-Hall effect (ISHE).~\cite{hahn2014, wang2014, qiu2015, lin2016} Similar results were also revealed in (Pt, Ta)/IrMn/CoFeB or Pt/NiO/FeNi heterostructures by spin-torque ferromagnetic resonance (ST-FMR) technique, where the spin current generated by spin-Hall effect (SHE) in Pt or Ta layer can propagate through IrMn or NiO layer and change the FMR linewidth.~\cite{mriyama2014,reichlova2015, mriyama2015} The spin current generated by spin pumping or spin Seebeck was also observed in IrMn/YIG, Cr/YIG, and $X$Mn/Py ($X$ = Fe, Pd, Ir, and Pt) bilayers through ISHE.~\cite{mendes2014, frangou2016, qu2015, zhang2014, sinova2015} Moreover, the IrMn/YIG, Pt/Cr$_2$O$_3$, and Pt/MnF$_2$ exhibit spin-Hall magnetoresistance (SMR) and large ISHE voltage, respectively, implying that the AFMs can be both spin-current detector and generator.~\cite{zhou2015, Seki2015, wu2016} These investigations open up new opportunities in developing the AFMs-based spin-current devices.

The IrMn alloy, which have been widely used to pin an adjacent FM layer in spin valve devices via exchange bias,~\cite{hoshino1996} demonstrates large ISHE voltage when in contacts with YIG.~\cite{mendes2014} Recently, a large SHE and anomalous-Hall effect (AHE) have been theoretically proposed in Cr, FeMn, and IrMn AFMs owing to their large spin-orbit coupling (SOC) or Berry phase of the non-collinear spin textures.~\cite{chen2014, shindou2001, freimuth2010} These theoretical predictions were also found to be valid for other cubic non-collinear AFMs, e.g., SnMn$_3$ and GeMn$_3$, where the calculations have been repeated with comparable results.~\cite{kubler2014}
The experimental investigation of AHE and SHE on the AFMs could be helpful from both fundamental and practical viewpoints for AFMs spintronics. As previously revealed in Cr/YIG bilayers, the large anomalous-Hall resistance (AHR) in thin unprotected Cr film is likely caused by the surface FM Cr oxides.~\cite{qu2015} Similar situation is expected in unprotected IrMn/YIG bilayers. Since the Pt/YIG bilayer is well studied,~\cite{wu2013} in this study, we choose the Pt as cap layer to protect the IrMn from oxidation to investigate the AHE and SHE of IrMn by measuring the spin transport properties in Pt/IrMn/YIG heterostructures.

\begin{figure}[tbp]
     \begin{center}
     \includegraphics[width=3.4in,keepaspectratio]{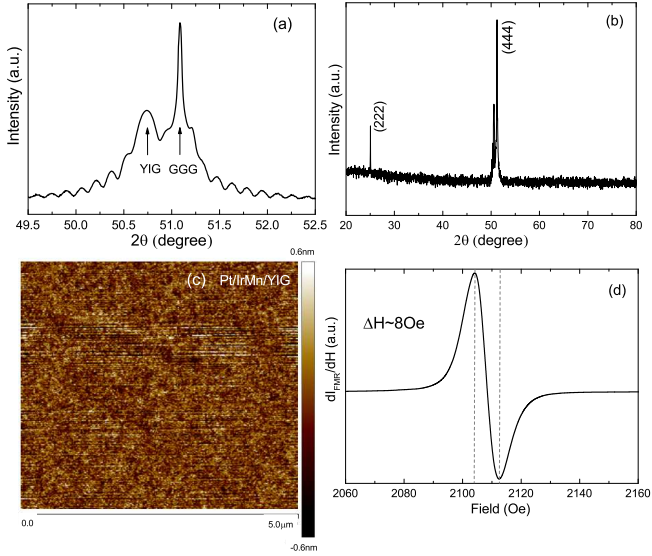}
     \end{center}
     \caption{(Color online) (a) A representative 2$\theta$-$\omega$ XRD patterns for YIG/GGG film near the (444) peaks of GGG substrate and YIG film. (b) The full range of XRD patterns from 20 to 80 degree. (c) Atomic force microscope surface topography of Pt/IrMn(3)/YIG heterostructure over an area of 5 $\mu$m $\times$ 5 $\mu$m. (d) A FMR derivative absorption spectrum of a 60 nm YIG film with an in-plane magnetic field; the line-width is estimated to be 8 Oe.}
     \label{fig1}
\end{figure}

\section{EXPERIMENTAL DETAILS}
The Pt/IrMn/YIG heterostructures were prepared in a combined ultra-high vacuum  (10$^{-9}$ Torr) pulsed laser deposition (PLD) and magnetron sputter system. The high quality epitaxial YIG films were deposited on (111)-orientated single crystalline Gd$_3$Ga$_5$O$_{12}$ (GGG) substrate via PLD technique as described elsewhere.~\cite{shang2015Rh} The Ir$_{20}$Mn$_{80}$ (IrMn) and Pt films were sputtered at room temperature in argon atmosphere in an $in$ $situ$ process. The thickness and crystal structure of films were characterized by Bruker D8 Discover high-resolution x-ray diffractometer (HRXRD). The thickness was estimated by using the software package LEPTOS (Bruker AXS). The surface topography of the films was measured in a Bruker Icon atomic force microscope. The ferromagnetic resonance (FMR) was measured by Bruker electron spin resonance spectrometers. The measurements of transverse Hall resistance, longitudinal resistance, and magnetization were carried out in a Quantum Design physical properties measurement system (PPMS) with a rotation option and magnetic properties measurement system (MPMS), respectively.

\section{RESULTS AND DISCUSSION}
Figure 1(a) plots a representative room-temperature 2$\theta$-$\omega$ XRD pattern of epitaxial YIG/GGG film near the (444) reflections. Clear Laue oscillations indicate the flatness and uniformity of the epitaxial YIG film. As shown in the Fig. 1(b), only the (222) and (444) reflections can be observed, and no indication of impurities or misorientation was detected in the full range of 2$\theta$-$\omega$ scan. In this study, the thicknesses of YIG and Pt films, determined by simulation of the x-ray reflectivity (XRR) spectra, are approximately 60 nm and 3 nm, respectively, while the IrMn thickness ranges from 0 nm to 8 nm. The atomic force microscope surface topography of Pt/IrMn(3)/YIG heterostructure over an area of 5 $\mu$m $\times$ 5 $\mu$m in Fig. 1(c) reveals a root-mean-square surface roughness of 0.18 nm, indicating atomical flat of prepared films. The other films show similar surface roughness. The number in the brackets represents the thickness of IrMn layer in nm unit. A representative FMR derivative absorption spectrum of YIG film (60 nm) shown in Fig. 1(d) exhibits a line width $\Delta$H = 8 Oe, which was measured at radio frequency 9.39 GHz and power 0.1 mW with an in-plane magnetic field at room temperature. The above properties indicate excellent quality of our prepared films.

\begin{figure}[tbp]
\begin{center}
\includegraphics[width=3.5in,keepaspectratio]{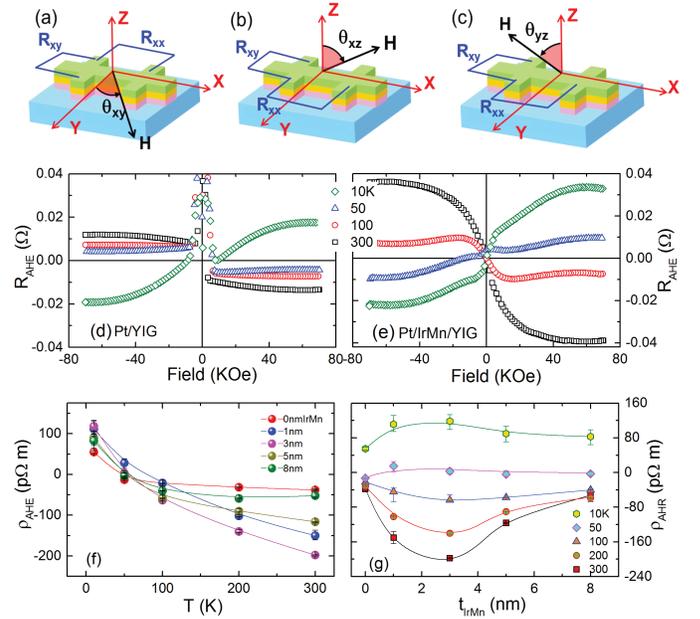}
\end{center}
\caption{(Color online) (a)-(c) Schematic plot of longitudinal resistance and transverse Hall resistance measurements. The magnetic fields are applied in the $xy$, $xz$, and $yz$ planes with angles $\theta_{xy}$, $\theta_{xz}$, and $\theta_{yz}$ relative to the $y$-, $z$-, and $z$-axes. The electric current is applied along the $x$-axis. Anomalous-Hall resistance R$_\textup{AHR}$ for Pt/YIG (d) and Pt/IrMn(1)/YIG (e) as a function of magnetic field at different temperatures. (f) Temperature dependence of the $\rho_\textup{AHR}$ for Pt/IrMn/YIG with various IrMn thicknesses.  The $\rho_\textup{AHR}$ are replotted as a function of IrMn thickness at various temperatures in (g). All $\rho_\textup{AHR}$ are averaged by [$\rho_\textup{AHR}$(70 kOe)- $\rho_\textup{AHR}$(-70 kOe)]/2. The error bars are the results of subtracting OHR in different field ranges}
\label{fig2}
\end{figure}

\begin{figure}[tbp]
     \begin{center}
     \includegraphics[width=3.4in,keepaspectratio]{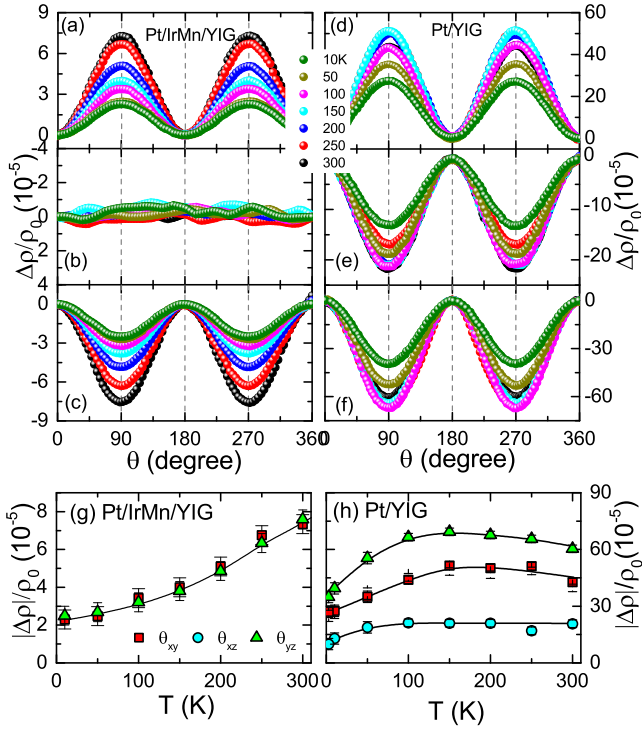}
     \end{center}
     \caption{(Color online) Anisotropic magnetoresistance for Pt/IrMn(1)/YIG at various temperatures down to 10 K with the magnetic field varied within $xy$ (a), $xz$ (b), and $yz$ (c) planes. The results of Pt/YIG are shown in (d)-(f). Temperature dependence of AMR amplitudes for Pt/IrMn(1)/YIG (g) and Pt/YIG (h) heterostructures. The cubic, circle and triangle symbols stand for the $\theta_{xy}$, $\theta_{xz}$, $\theta_{yz}$ scans, respectively.}
     \label{fig3}
\end{figure}

\subsection{Anomalous-Hall resistance}

As shown in the top panel of Fig. 2, in order to measure the transverse Hall resistance and longitudinal resistance, all the Pt/IrMn/YIG heterostructures were patterned into Hall-bar configuration (central area: 0.3 mm $\times$ 10 mm; electrode 0.3 mm $\times$ 1 mm). The transverse Hall resistance R$_{xy}$ of Pt/IrMn/YIG was measured in the temperature range of 10 K to 300 K with perpendicular magnetic field ranging from -70 to 70 kOe. In metal thin film, the ordinary-Hall resistance (OHR) R$_\textup{OHR}$ is subtracted from the measured R$_{xy}$, i.e., R$_\textup{AHR}$ = R$_{xy}$ - R$_\textup{OHR}$$\times \mu_0H$, where R$_\textup{AHR}$ is AHR. As shown in Figs. 2(d)-(e), the resulting R$_\textup{AHR}$ as a function of magnetic field for Pt/YIG and Pt/IrMn(1)/YIG are presented. It is noted that the Pt becomes magnetic when in contacts with the YIG due to its proximity to the stoner ferromagnetic instability, i.e., magnetic proximity effect (MPE), as previously shown experimentally by x-ray magnetic circular dichroism (XMCD) and theoretically by first-principles calculation.~\cite{liu2013, qu2013} The magnetized Pt shares some common features as magnetic YIG film, i.e., strong anisotropy.~\cite{shang2015Rh} Thus, when the magnetic field approaches zero, the magnetized Pt moments are randomly distributed, the R$_\textup{AHR}$ exhibits irregular M-shaped behavior close to zero field. However, for Pt/IrMn/YIG, the R$_\textup{AHR}$ continuously decreases as approaching zero field, implying that the Pt/IrMn and IrMn/YIG interfaces are free of MPE, being consistent with the absence of conventional anisotropic magnetoresistance (CAMR) (see below). We summarize the derived anomalous-Hall resistivity $\rho_\textup{AHR}$ of Pt/IrMn/YIG heterostrutures as functions of temperature ($T$) and IrMn thickness ($t_\textup{IrMn}$) in Figs. 2(f)-(g). The $\rho_\textup{AHR}$($T$) for all the Pt/IrMn/YIG exhibits rich characteristics whose magnitude and sign are highly non-trivial, which were also found in Pt/LaCoO$_3$ bilayers.~\cite{shangLCO}. As shown in Fig. 2(h), the magnitude of $\rho_\textup{AHR}$ decrease with temperature and then it increases again below 100 K. Simultaneously, the $\rho_\textup{AHR}$ change its sign at the temperature which is independent of IrMn thickness. We also replotted all the $\rho_\textup{AHR}$ as a function of IrMn thickness in Fig. 2(g). In the studied temperature range, as increasing the $t_\textup{IrMn}$, the $\rho_\textup{AHR}$ also increases and reaches a maximum around $t_\textup{IrMn}$ = 3 nm, which excludes the interfacial origin of the observed AHR.

\subsection{Spin-Hall magnetoresistance}
The anisotropic magnetoresistance (AMR) for Pt/IrMn/YIG was also measured down to low temperatures. As an example, the AMR of Pt/IrMn(1)/YIG and Pt/YIG for three different field scans are presented in top panel of Fig. 3. When the magnetic field scans within the $xy$ plane [Fig. 3(a)(d)], both the CAMR and SMR contribute to the total AMR; for the $xz$ plane [Fig. 3(b)(e)], the resistance changes are attributed to the MPE-induced CAMR; for the $yz$ plane [Fig. 3(c)(f)], the CAMR is zero, and only the SMR are expected.~\cite{chen2013,isasa2014} As shown in Fig. 3(b), the $\theta_{xz}$ scan shows negligible AMR and the resistance is almost independent of $\theta_{xz}$, indicating the extremely weak MPE at the interface even down to low temperatures. However, the MPE is significant at Pt/YIG interface [see Fig. 3(e)]: the maximum amplitude of CAMR is around 2.2 $\times$ 10$^{-4}$, which is comparable to the SMR. Thus, the IrMn can be used as clean spin current detector and generator, similar to the normal Rh or AFM Cr metals.~\cite{shang2015Rh, qu2015} Since the CAMR is negligible in Pt/IrMn(1)/YIG, the SMR dominates the AMR when the magnetic field is varied within the $xy$ plane, the amplitudes of $\theta_{xy}$ scan are almost identical to $\theta_{yz}$ scan. While for Pt/YIG, due to the MPE-induced CAMR, none of the amplitudes is identical to each other. The temperature dependence of the AMR amplitudes for all $\theta_{xy}$, $\theta_{xz}$, and $\theta_{yz}$ scans are summarized in Fig. 3(g) and Fig. 3(h) for Pt/IrMn(1)/YIG and Pt/YIG, respectively. Upon decreasing the temperature, the SMR persists down to 10 K, with the amplitudes monotonically decreasing from 7.5 $\times$ 10$^{-5}$ (300 K) to 3.0 $\times$ 10$^{-5}$ (10 K) in Pt/IrMn(1)/YIG. For Pt/IrMn/YIG, the amplitudes of SMR are almost an order smaller than that of the Pt/YIG due to the smaller spin-Hall angle, shorter spin diffusion length, and larger electrical resistivity of IrMn.~\cite{chen2013, mendes2014, zhang2014} The temperature characteristics of SMR amplitudes in Pt/IrMn/YIG are significantly different from the Pt/YIG or Pd/YIG bilayers, where the SMR amplitudes exhibit nonmonotonic temperature dependence and acquire a maximum around 100 K.~\cite{marmion2014, lin2014} For Pt/YIG, the temperature dependence of SMR amplitude can be described by a single spin-relaxation mechanism.~\cite{marmion2014} The spin diffusion length is defined as $\lambda = \sqrt{D \tau_{sf}}$, where $D$ and $\tau_{sf}$ are diffusion constant and spin-flip relaxation time, respectively. Within the Elliot-Yafet spin-orbit scattering model, both $D$ and $\tau_{sf}$ are proportional to the reciprocal of temperature dependence of the resistivity 1/$\rho(T)$.~\cite{Fabian1998,liu2012} In Pt metal, the electrical resistivity mainly comes from phonon-electron scattering at high temperature, then $\lambda \propto 1/T$. However, the extra magnetic electron scattering need to be considered in Pt/IrMn/YIG heterostructures, the assumption of $\lambda \propto 1/T$ is invalid. It is noted that the heterostructures with different IrMn thicknesses exhibit similar temperature dependent characteristics with different numerical values compared to the Pt/IrMn(1)/YIG heterostructure shown here. For example, the Pt/IrMn(3)/YIG exhibits the SMR amplitude of 6.8 $\times$ 10$^{-5}$ at room temperature. The sizable SMR observed in Pt/IrMn/YIG heterostructures indicates that the spin current can transport through IrMn layer.

\begin{figure}[tbp]
     \begin{center}
     \includegraphics[width=3.4in,keepaspectratio]{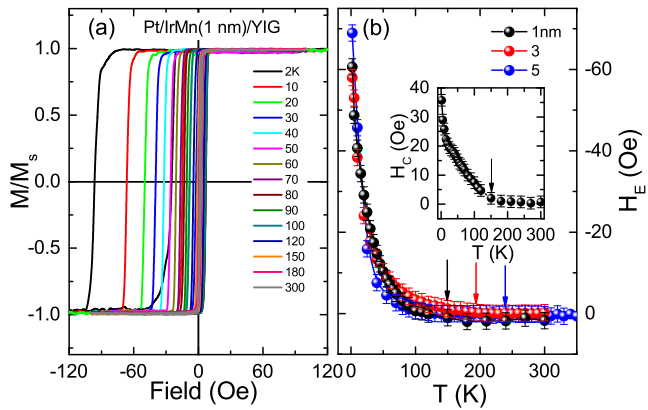}
     \end{center}
     \caption{(Color online) (a) Field dependence of normalized magnetization $M$/$M_\textup{s}$ for Pt/IrMn(1)/YIG at various temperatures down to 2 K. The magnetic field is applied parallel to the film surface. The paramagnetic background of the GGG substrate has been subtracted. (b) The in-plane exchange bias field $H_\textup{E}$ versus temperature. The arrows indicate the AFM block temperatures $T_\textup{b}$. The inset plots the coercivity field $H_\textup{C}$ as a function of temperature for Pt/IrMn(1)/YIG.}
     \label{fig4}
\end{figure}

\subsection{Magnetization}
Since the magnetic transitions of very thin AFMs are expected to be well below the ordering temperature of bulk forms, we measured the field dependence of magnetization down to low temperatures, from which we can track the AFM blocking temperature $T_\textup{b}$ for Pt/IrMn/YIG heterostructures. As an example, the normalized magnetic hysteresis loops $M$/$M_\textup{s}$ for Pt/IrMn(1)/YIG at various temperatures after field cooling from 300 K are presented in Fig. 4(a). The derived exchange bias field $H_\textup{E}$ versus temperature are summarized in Fig. 4(b), from which the $T_\textup{b}$ are approximately estimated to be 150 K, 180 K, and 220 K for 1 nm, 3 nm and 5 nm IrMn, respectively, as the arrow indicated. Similar blocking temperatures were previously reported in IrMn/MgO/Ta tunnel junctions and IrMn/NiFe bilayer.~\cite{petti2013, tshitoyan2015}  Moreover, the coercivity $H_\textup{C}$ also exhibits a step-like increase near the blocking temperature, as the arrow shown in the inset of Fig. 4(b), indicating the strongly enhanced exchange coupling between IrMn and YIG layer below $T_\textup{b}$.

\subsection{Discussion}
Based on the above experimental results, we discuss the origins of the significant AHR in Pt/IrMn/YIG heterostructures and the effect of AFM order on spin transport properties. There are at least four contributions to the observed AHR in Pt/IrMn/YIG: MPE, spin-Hall based SMR, spin-dependent interface scattering, and intrinsic properties of IrMn metal. In contrast to the Pt/YIG, the negligible CAMR in Pt/IrMn/YIG indicates the extremely weak MPE at Pt/IrMn or IrMn/YIG interfaces, which is different from the previous studied of IrMn/YIG bilayer.~\cite{zhou2015} The SMR model based on SHE also predicts an anomalous-Hall-like resistance,~\cite{chen2013} whose magnitude and sign are determined by the spin diffusion length and spin-Hall angle of the metal and the imaginary part of the spin mixing conductance, respectively. Though the thickness dependence of the AHR in Pt/IrMn/YIG can be described by the SMR model, it fails to explain the AHR by the following reasons: (i) An arbitrary temperature dependence of the imaginary part of the spin mixing conductance parameter is required to qualitatively describe the temperature-dependent AHR data, i.e., sign reversal; (ii) According to the spin pumping studies, both the spin-Hall angle and the spin diffusion length of IrMn are smaller than Pt, which cannot explain the enhancement of AHR by increasing the IrMn thickness.~\cite{mendes2014, zhang2014, sinova2015} Spin-dependent scattering at the interface, combined with the conventional skew-scattering and side-jump mechanisms, can also give rise to AHR.~\cite{nagosa2010} Again, the enhancement of AHR by increasing the IrMn thickness excludes the interfacial origin. Finally, the theoretical calculations predict a large AHE and SHE in IrMn metal not only attributed to the large SOC of heavy Ir atoms which is transferred to the magnetic Mn atoms by hybridization effect but also the Berry phase of the non-collinear spin structures.~\cite{ chen2014,shindou2001, freimuth2010} We conclude that the large AHR observed in Pt/IrMn/YIG is likely associated with SOC and non-collinear magnetic structure of IrMn. However, the non-trivial temperature dependence of AHR demands further theoretical and experimental investigations.

Now we discuss the possible interplay between AFM order and spin transport properties. As shown in Fig. 2 and Fig. 3, there is no clear anomalous in AHR or SMR near the blocking temperatures of IrMn, implying weak correlations between the AHE or SHE and the AFM order in IrMn. Similar results were also observed in Cr/YIG bilayers, where the ISHE voltage and AHR is also independent of AFM ordering temperature.~\cite{qu2015} According to our magnetization results (Fig. 4), the AFM ordering temperatures of our IrMn films are well below room temperature. However, the enhancement of AHR in Pt/IrMn/YIG happens in the whole studied temperature range [see Fig. 2(g)]. There are two possible reasons for this phenomenon, one is that the AHE and SHE attributed to non-collinear magnetism is generated on a length scale of nanometer and is a local property not relying on long range magnetic order, i.e., regardless of how IrMn grains are orientated, as reported previously in Mn$_5$Si$_3$ film.~\cite{surgers2016} The second one is that the strength of SOC is independent of AFM order in IrMn metal, which is mainly determined by the Ir atoms.

\section{CONCLUSIONS}
In summary, we report an investigation of AHE and SHE by measuring the AHR and SMR in Pt/IrMn/YIG heterostrucutres. The significant AHR in Pt/IrMn/YIG is likely associated with the strong SOC and non-collinear magnetic structure of IrMn, and the sizable SMR indicates that the spin current can transport through IrMn. The observed non-trivial temperature dependence of AHR cannot be consistently explained by the existing theories, further investigations are needed to clarify this issue. Moreover, both the AHR and SMR are uncoupled to the AFM order of IrMn metal. The negligible MPE at Pt/IrMn or IrMn/YIG interface and large ISHE indicate that IrMn can be another model system to explore physics and devices associated with antiferromagnetism and pure spin current.

\begin{acknowledgments}
We thank the high magnetic field laboratory of Chinese Academy of Sciences for the FMR measurements. This work is financially supported by the National Natural Science foundation of China (Grants No. 11274321, No. 11404349, No. 51502314, No. 51522105) and the Key Research Program of the Chinese Academy of Sciences (Grant No. KJZD-EW-M05). S. Zhang was partially supported by the U. S. National Science Foundation (Grant No. ECCS-1404542).
\end{acknowledgments}

\end{document}